\documentclass[10pt, conference, compsocconf]{IEEEtran}
\makeatletter

\def\ps@IEEEtitlepagestyle{%
    \def\@oddfoot{\mycopyrightnotice}%
    \def\@evenfoot{}%
}
\def\mycopyrightnotice{%
    {\footnotesize  2325-2944/19/\$31.00 \textcopyright2019 IEEE\hfill} \newline {\footnotesize DOI 10.1109/DCOSS.2019.00042\hfill}
    \gdef\mycopyrightnotice{}
}

\makeatletter
\newcommand*\titleheader[1]{\gdef\@titleheader{#1}}
\AtBeginDocument{%
  \let\st@red@title\@title
  \def\@title{%
    \bgroup\normalfont\large\centering\@titleheader\par\egroup
    \vskip1.5em\st@red@title}
}
\makeatother

\usepackage{ifpdf}
 \ifpdf
 \else
 \fi

\usepackage{cite}

\ifCLASSINFOpdf
  \usepackage[pdftex]{graphicx}
  \DeclareGraphicsExtensions{.pdf,.jpeg,.png,.PNG}
\else
  \usepackage[dvips]{graphicx}
  \DeclareGraphicsExtensions{.eps}
\fi

\usepackage{natbib}
\usepackage[cmex10]{amsmath}
\usepackage{algorithmic}
\usepackage{array}
\usepackage{mdwmath}
\usepackage{mdwtab}
\usepackage{eqparbox}
\usepackage[font=footnotesize]{subfig}
\usepackage{fixltx2e}
\usepackage{stfloats}
\usepackage{url}

\title{A Model for Reliable Uplink Transmissions in LoRaWAN}
\titleheader{2019 15th International Conference on Distributed Computing in Sensor Systems (DCOSS)}

\begin{document}
%

\author{\IEEEauthorblockN{Furqan Hameed Khan}
\IEEEauthorblockA{\textit{School of ITEE} \\
\textit{The University of Queensland}\\
Brisbane, Australia \\
furqan.khan@uq.net.au}
\and
\IEEEauthorblockN{Raja Jurdak}
\IEEEauthorblockA{\textit{Distributed Sensing Systems} \\
\textit{Data61 CSIRO}\\
Brisbane, Australia \\
rjurdak@ieee.org}
\and
\IEEEauthorblockN{Marius Portmann}
\IEEEauthorblockA{\textit{School of ITEE} \\
\textit{The Univeristy of Queensland}\\
Brisbane, Australia \\
marius@ieee.org}
}


\maketitle

\begin{abstract}
Long range wide area networks (LoRaWAN) technology provides a simple solution to enable low-cost services for low power internet-of-things (IoT) networks in various applications. The current evaluation of LoRaWAN networks relies on simulations or early testing, which are typically time-consuming and prevent effective exploration of the design space. This paper proposes an analytical model to calculate the delay and energy consumed for reliable Uplink (UL) data delivery in Class A LoRaWAN. The analytical model is evaluated using a real network test-bed as well as simulation experiments based on the ns-3 LoRaWAN module. The resulting comparison confirms that the model accurately estimates the delay and energy consumed in the considered environment. The value of the model is demonstrated via its application to evaluate the impact of the number of end-devices and the maximum number of data frame retransmissions on delay and energy consumed for the confirmed UL data delivery in LoRaWAN networks. The model can be used to optimize different transmission parameters in future LoRaWAN networks.
\end{abstract}


%
\IEEEpeerreviewmaketitle

\section{Introduction}
\label{sec:intro}
Future 5G networks will support applications that require ultra-reliable communication among massive Internet-of-things (IoT) devices~\cite{IEEEhowto:00}. Achieving low-cost and reliable communication with bounded delay is challenging in these networks. Moreover, the number of IoT devices in future ubiquitous networks will eventually surpass the number of connected user devices~\cite{IEEEhowto:01}, making it more difficult to satisfy the application-specific requirements of data rate and latency. 
\begin{figure}[!t]
\begin{center}
\resizebox{3.3 in}{1.8 in}{\includegraphics{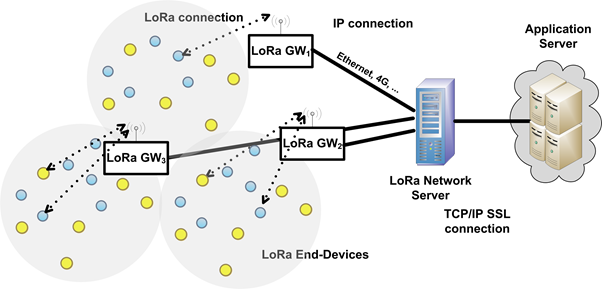}}
\caption{\em Long Range Wide Area Network (LoRaWAN) Architecture}
\label{jpg1}
\end{center}
\end{figure}

A new suite of long-range IoT technologies have emerged, including long term evolution for machines (LTE-M), narrow-band IoT (NBIoT), SigFox, Ingenu, and long range (LoRa)~\cite{IEEEhowto:22}. These technologies target low rates, small power, and wide area coverage for a massive number of devices. Among these, LoRa provides low cost, low power, and long range IoT solution, targeting various network devices such as wearables, smart meters, and remote sensors. This technology uses the LoRa modulation scheme over its physical layer. LoRa radio transmission requires setting up of the following five configurable parameters with respective limits: transmission power (2 dBm to 20 dBm), frequency (860 MHz to 1020 MHz), spreading factor (SF) (7 to 12), channel bandwidth (125kHz, 250 kHz, or 500 kHz), and coding rate (4/5 to 4/8)~\cite{IEEEhowto:18}. 

LoRaWAN is the medium access layer (MAC) protocol used by LoRa devices to transmit their data over the wireless channel. So far, LoRaWANv1.1 has defined the operation of three device classes, namely Class A (sensors), Class B (actuators), and Class C (all time active) in a LoRa network~\cite{IEEEhowto:11}. Class A and B devices perform energy efficient network operation in the uplink (UL) and downlink (DL) respectively. All LoRaWAN devices must support the Class A UL operation, which forms the basis of LoRaWAN operation and is the focus of this work. Fig.~\ref{jpg1} presents the LoRaWAN network architecture consisting of end devices, gateways, and a network server. The network server is connected to the end devices over the MAC layer running LoRaWAN, whereas the LoRa gateways are just traffic forwarding elements. The server is the main component of the LoRa network that handles traffic coming via all gateways. LoRaWAN devices support two different types of frames transmissions, namely \emph{confirmed} and \emph{unconfirmed}. The confirmed frames require an acknowledgement (ACK) back to the end device, whereas the unconfirmed frames do not need any ACK. This work focuses on the UL communication through the confirmed frames transmissions. 

Achieving reliable and low-cost data delivery in future IoT networks is important. Current practice in exploring the design space for LoRa networks mostly relies on simulations, with some early testing efforts. Simulations can be time-consuming and limiting due to their iterative nature, while physical experimentation is costly, tedious, and labour-intensive. Thus, there is a clear need for a mathematical model to more effectively explore the design space and highlight performance tradeoffs in different scenarios.

We propose an analytical model to calculate the amount of resources consumed for the UL delivery of confirmed data frames. Empirical experiments based on a small scale LoRaWAN testbed, as well as larger scale ns-3 simulations, are used to evaluate our model in terms of the resources consumed by each end-device for an UL confirmed transmission. Our findings confirm that the proposed model can be used to estimate the consumed energy and delay experienced by an end device for reliable UL transmission. Next, we further evaluate our Markov-chain based analytical model, by exploring different values for the maximum number of allowed retransmissions, the choice of receive slot for ACK transmissions, and the network load. Our model shows how transmission delay and energy depend on these parameters.

In the rest of this paper, the next section provides a background on the basic LoRaWAN operation. Section III discusses the model assumptions and explains the complete Markov-chain based model and its associated state transition probabilities. In Section IV we show the results attained through our proposed model, from a small scale test-bed, as well as using the ns-3 LoRaWAN module based simulations. Section V presents and discusses the evaluation results of different use case scenarios based on our proposed analytical model. Section VI briefly outlines the relevant related work, and Section VII concludes the paper. 

\section{Background}
\label{sec:back}
\begin{figure}[t!]
\begin{center}
\resizebox{3.2 in}{3.8 in}{\includegraphics{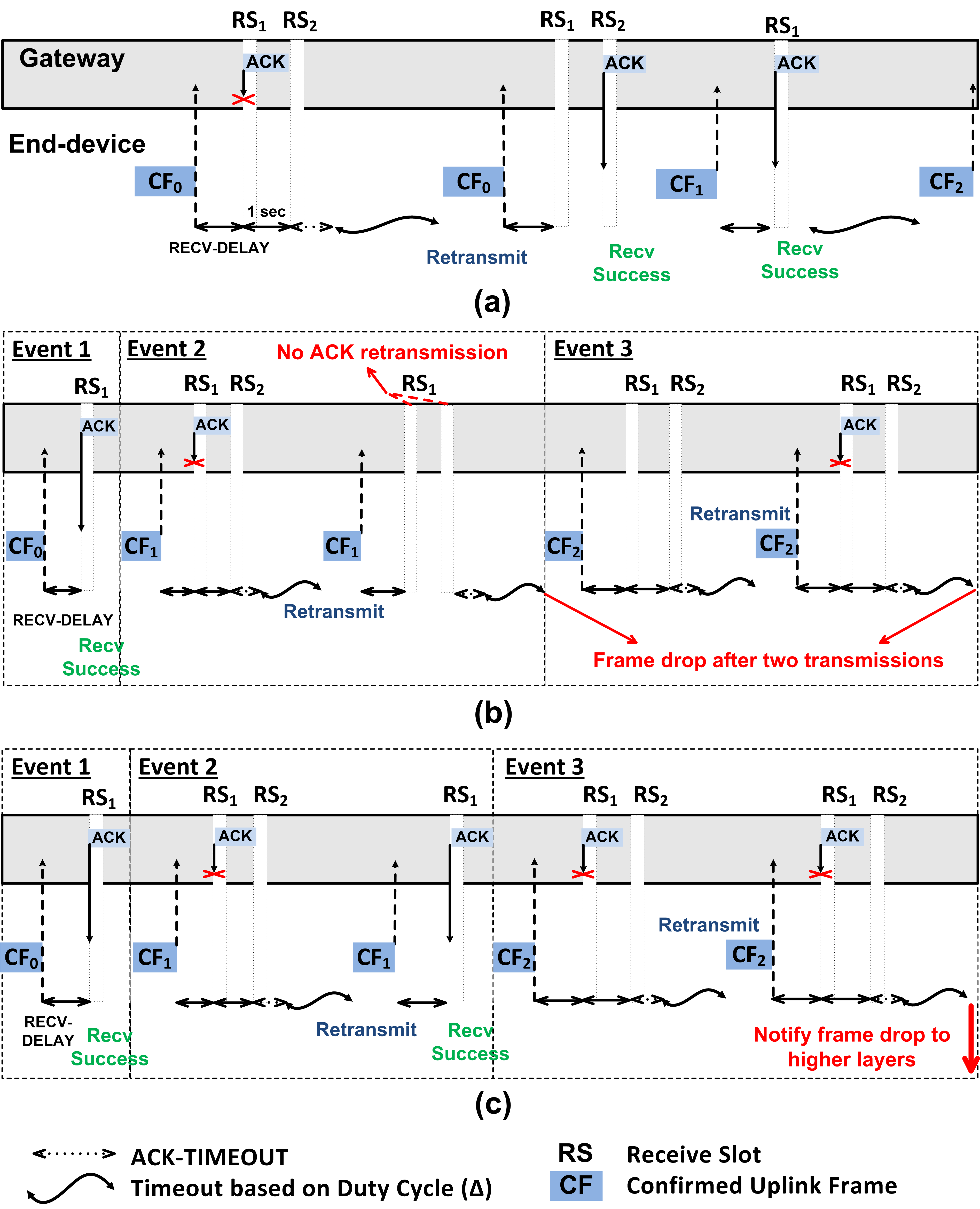}}
\caption{\em Confirmed UL Transmission in LoRaWANv1.1: (a) General scenario, (b) Case 1, and (c) Case 2 for confirmed UL frame with maximum transmissions N = 2. }
\label{jpg2}
\end{center}
\end{figure}
This section briefly describes the basic working of LoRaWAN. In LoRaWANv1.1, a Class A device performs UL transmission following the pure ALOHA access method, while respecting the region-specific regulatory duty cycle (RDC) restriction over the specific sub-band. Various sub-bands are supported in LoRaWAN, some consist of only main channels and are used for normal data transmission, some are used by devices to send join requests, while others are reserved for sending DL responses from the gateway. For each UL transmission, the device switches to transmit mode and picks a spreading factor (SF), channel, bandwidth, coding rate (as advised from the network server), and sends the data for the transmit duration based on the frame length. After the UL transmission, the end device opens two consecutive receive slots as shown in Fig.~\ref{jpg2}a, to wait for DL data transmission. The two receive slots recommends offset value of 1 and 2 seconds respectively. As shown in Fig.~\ref{jpg2}a, after the~\emph{receive delay} expiration, in receive slot 1 ($RS_1$) the device listens for any DL frames over the same channel with a rate which is a function of the UL data rate. If it hears nothing, the device wait for another second and again listens for the DL data in receive slot 2 ($RS_2$) over a reserved channel with minimum rate. If the device does not get any ACK during $RS_2$ duration (equal to preamble air-time with minimum rate), it waits for an ACK-TIMEOUT duration, to make sure that the device has nothing to send until the air-time of a valid frame. After the time expires, the same confirmed frame $CF_{0}$ (with frame counter zero) can be retransmitted following the RDC, similar to any other UL transmission, as shown in the case of $CF_{0}$ in Fig.~\ref{jpg2}a. From Fig.~\ref{jpg2}a, note that the LoRaWAN operation results in two transmission opportunities from the gateway as a result of an UL transmission from a device. This is to save the battery powered devices resources, in contrast to the gateway which is often mains powered, or has otherwise a less restricted power source.

A device will continue to re-transmit frame following the RDC, as long as it does not get any DL data from the gateway. The maximum number of confirmed frame retransmissions is a configurable value, with a recommended limit of seven\footnote{i.e. if N is the maximum number of retransmissions including the first transmission, then N $=$ 8 as defined in the LoRaWANv1.1~\cite{IEEEhowto:11}.}. The device moves to the next frame once the current frame gets acknowledged, or the limit $N$ is reached. Furthermore, for each received data frame (with unique frame counter), the gateway must send the ACK frame only once~\cite{IEEEhowto:11}, as seen from the operation depicted in Fig.~\ref{jpg2}b, i.e. Case 1. However, to enhance the network performance, most LoRaWAN applications support re-sending of ACK frames, as shown by the process in Fig.~\ref{jpg2}c, i.e. Case 2. Moreover, Fig.~\ref{jpg2}b,c show three different conditions of ACK transmission in the case of a confirmed data frame (with $N$ = 2). In the first, the ACK frame is immediately received by the end device in $RS_1$. In the second case, the ACK frame is lost during the first transmission attempt, whereas in the last one the frame could not be successfully ACK'd. If after all the $N$ attempts no ACK for the confirmed frame has arrived, the device will drop the frame immediately after the ACK timeout. Note further in Event 3 of Fig.~\ref{jpg2}b,c,  after the UL frame reception in the first transmission attempt, the gateway can send the ACK in one of the two receive slots. After the ACK from the gateway is lost, the device will assume that the transmission was not successful and will resend the frame after the ACK-TIMEOUT. Finally, when the device PHY layer does not get any ACK after $N$ attempts, it will drop the frame and will notify the loss to the higher layer. 
\section{System Model}
%
%
\begin{figure*}[t!]
\begin{center}
\resizebox{5.5 in}{2.4 in}{\includegraphics{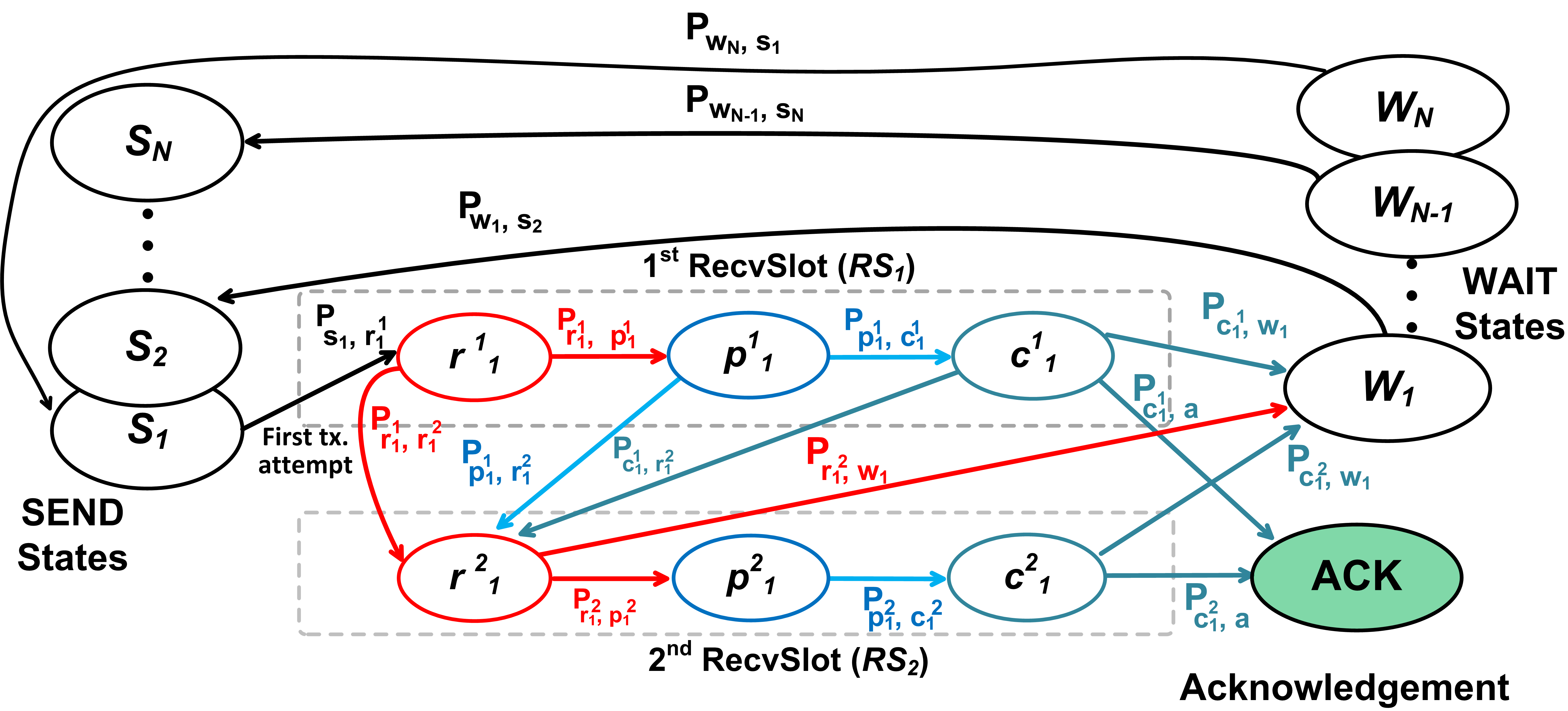}}
\caption{\em Proposed Markov-Chain Model for Confirmed UL Transmission in LoRaWAN.}
\label{jpg3}
\end{center}
\end{figure*}
In this section, we develop an analytical model for the UL confirmed data frame transmission of class A devices in a LoRaWAN network. For a given traffic load over a device and in the presence of a given number of devices, our goal is to estimate the average time it takes for the successful reception of a data frame sent by an arbitrary end device. Also given a maximum transmissions limit, our model can be used to find the worst case delay and energy consumption. The notation and symbols used in this section are listed with their definitions in Table~\ref{table1:symbolsmeaning}.
\subsection{Assumptions}
We make the following assumptions for our model,
\begin{enumerate}
\item Only class A type LoRaWAN devices are present. Hence, an ACK does not contain any piggybacked data and has a fixed size comprising the preamble and the MAC frame header with a set ACK flag.
\item For the proposed analytical model, we consider that overlapping transmissions from multiple devices over the same channel with the same LoRaWAN transmission parameters (i.e. bandwidth, SF, code rate) will be lost. In other words, we ignore the capture effect.
\item All UL transmissions respect the RDC of the given sub-band. For $RS_2$, our model assumes that the reserved channels used for DL ACK transmission by the gateway has no duty cycle restriction. This is valid~\footnote{ETSI regulations~\cite{IEEEhowto:21} allows a device to use channel either by respecting the RDC or following listen before talk adaptive frequency agility.}, as only the gateway can use these reserved channels.
\item All network devices have uniform traffic intensities.
\end{enumerate}
\subsection{Proposed Analytical Model}
Fig.~\ref{jpg3} shows our proposed Markov-chain model for the confirmed UL transmission, where devices perform data transmission based on the region specific RDC restrictions~\cite{IEEEhowto:12}, and the data frame availability. In our model, for each new data frame all available channels have a uniform probability to be selected for transmission. Whenever an end-device PHY has a frame ready to send, it switches to the~\emph{Send Frame} state $S_1$ (where n=1, for the first transmission attempt of a confirmed UL data frame.), and remains there until the end of the frame transmission. Then the device wait in the receive slot $RS_1$ in state ($R^1$, n) to detect DL data over the same channel with the same rate that was used for the UL frame transmission. As shown in Fig.~\ref{jpg3}~\footnote{The arrows for transition between states with only non-zero state transition probabilities are shown in the Fig.~\ref{jpg3}.}, the states for the data reception in the first receive slot, i.e. ($R^1$, n), ($P^1$, n), and ($C^1$, n) are defined by $r^1_n$, $p^1_n$, and $c^1_n$. These are referred to as the detection of preamble ($r^1_n$), checking the preamble ($p^1_n$), and checking the ACK data frame ($c^1_n$), in $RS_1$, respectively. Note that here, $n$ refers to the $n^{th}$ transmission attempt of the given confirmed UL frame. Thus, n=1 is the first transmission, while n=2, 3, ... , N represent the later UL retransmissions of the same confirmed frame, until the ACK is received. Similar to $RS_1$, for every successive UL transmission, $RS_2$ also has three 2-dimensional states for checking the preamble and the corresponding frame data, namely ($R^2$, n), ($P^2$, n), and ($C^2$, n), represented by $r^2_n$, $p^2_n$, and $c^2_n$, as shown in Fig.~\ref{jpg3}.
\begin{table}[t!]
\caption{Symbols and their Definition}\centering
\label{table1:symbolsmeaning}
\begin{tabular}{|c|c|}
\hline
Symbols   & Definition \\
\hline
$A$ &Total network devices\\
\hline
$n$=1,2,...,$N$ &($n$) frame tx., ($N$) max. possible tx. \\
\hline
$t_I$ &Traffic intensity \\
\hline
$\Delta$ &Duty cycle of sub-band for UL tx. \\
\hline
$m_c$ &Channels available for UL tx. \\
\hline
$\alpha$ &Channel quality\\
\hline
$\gamma_n$ &Slot use by gateway for ACK tx.\\
\hline
$P_{s_n, r^1_n}$ &Transition probability from $s_n$ to $r^1_n$\\
\hline
$x_a(/y_a)$ &Probability of device $a$ tx. (/not tx.)\\
\hline
$D_{s_n}$ &Delay observed in state $s_n$\\
\hline
$E_{s_n}$ &Energy consumed in state $s_n$\\
\hline
$t^{TX}_{n}$, $t^{PR}_{n}$, $t^{ACK}_{n}$&Data, preamble, ACK frames air-time\\
\hline
\end{tabular}
\end{table}

\textbf{\underline{Model Description}:} We assume a LoRaWAN network with available sub-bands, comprising of $m_{c}$ channels available for UL data transmission. $\Delta$ is the maximum percentage of air-time a device can use sub-bands for data frame transmission, and it is called as the RDC constraint. With this restriction, a device can only transmit a constant number of fixed size UL packets to the gateway. We call it the packet transmission rate of an end-device. If the frame arrival rate to an end-device application is greater than or equal to its frame transmission rate (regulated by $\Delta$), the network is considered to be fully loaded. We use $t_I$ to represent the ratio of the frame arrival rate and frame transmission rate from a particular device. In other words, $t_I$ ($0$ $\leq$ $t_I$ $\leq$ $1$) defines the traffic intensity. From here on, we use $t_I$ $=$ $1$, i.e. all devices in the network are fully loaded. Also, as seen from Table~\ref{table1:symbolsmeaning}, $\gamma_n$ is the choice of receive slot used by gateway for sending ACK in response to the $n^{th}$ data frame transmission. $\gamma_n$ if one and zero implies that the $RS_{1}$ and $RS_{2}$ is chosen, respectively.

The overall traffic generation probability over the sub-band is given by $\Delta$ $\cdot$ $A$. For a device $a$, the probability of sending frame within the duty cycle over a channel will be,
\begin{equation}\label{eq03}
	x_a = \frac{\Delta}{m_c},
\end{equation}
As in~\cite{IEEEhowto:06}, the probability that an active device will not send data over a channel is defined through $y_a$ as 1 - $\frac{\Delta}{m_c}$.
%
%
From LoRaWAN specifications~\cite{IEEEhowto:11}, during retransmission the device must not use the same channel that was used in the last transmission. Hence with $m_c$ available channels for the first transmission, the available channels will reduce by one in the following retransmission. Therefore the probability that an end device is not using a given channel during retransmission is,
\begin{equation}\label{eq05}
	y'_a = 1 - \frac{\Delta}{m_c - 1},
\end{equation}

The formulation of the state transition matrix $\mathbf{P}$ for our model is added in the Appendix. A detailed version with all relevant explanation is available at~\cite{IEEEhowto:24}.

To find the expected delay and energy consumed using the Markov-chain model of Fig.~\ref{jpg3}, we use the resulting state transition matrix $\mathbf{P}$ of length (8 $\cdot$ N + 1) $\times$ (8 $\cdot$ N + 1). Using $J^{0}$ as the row vector of length 1 $\times$ (8 $\cdot$ N + 1), with initial value for the state $S_1$ as 1. The probability of a device residing in a particular state can be determined after each K period as, $J^{(K)}$ $=$ $J^{0}$ $\cdot$ ${\mathbf{P}}^{K}$. Using the Markov-chain property as $K$ approaches to infinity, the probabilities for each state in $J^{(K)}$ approaches to some steady state. Based on these steady state probabilities ($J^{(K)}$ for very large $K$) we can find the average delay and energy consumed as a device moves from the send state ($S_1$) to the $ACK$ state, using the delay and energy values for each state.

\textbf{\underline{Delay vector}:} The delay in each state is given by the column vector $\mathbf{D}$ (of length 8 $\cdot$ N + 1) defined by [$D_{s_n}$, $D_{r^1_n}$, $D_{p^1_n}$, $D_{c^1_n}$, $D_{r^2_n}$,  $D_{p^2_n}$, $D_{c^2_n}$, $D_{w_n}$, $D_a$], where $n$ corresponds to the $n^{th}$ transmission of the current data frame. Here, $D_{s_n}$ comprises the delay due to the $n^{th}$ data frame transmission by the user device until the time the end device waits for $RS_1$ (1 + $t^{TX}_{n}$). $D_{r^1_n}$ is the time a device spends in $RS_1$ to receive/decode the DL preamble during $n^{th}$ frame transmission ($t^{PR}_{n}$). $D_{p^1_n}$ is the time delay for checking the received preamble and it is assumed as negligible ($\approx$ 0). $D_{c^1_n}$ is the delay in receiving the entire DL ACK frame of 12 bytes ($t^{ACK}_{n}$). It has two components, the first component corresponds to the case when the UL frame gets ACK'd successfully in the state $RS_1$. The second is the case when the frame gets lost due to poor channel quality (low $\alpha$) and the device wait for 1 - $t^{ACK}_{n}$ seconds until it gets to $RS_2$. Thus $D_{c^1_n}$ for n=1 will be,
\begin{equation}\label{eq19}
	D_{c^1_1} = \gamma_1 {y_a}^{2 A} \alpha \cdot ((t^{ACK}_{1} - t^{PR}_{1}) + max(1 - t^{ACK}_{1}, 0)),
\end{equation}

Note that, for n $\geq$ 2 the variable $y_a$ will be replaced by $y'_a$. Similarly, the states ($R^2$, n), ($P^2$, n), and ($C^2$, n) are the same states for receiving and  preamble checking, and the payload data corresponds to $RS_2$. The delay for receiving the ACK preamble in $RS_2$, i.e. state ($R^2$, n), is the same as for $t^{PR}_{n}$, $D_{p^2_n}$ and $D_a$ are very small ($\approx$ 0), and $D_{c^2_n}$ is $t^{ACK}_{n}$ - $t^{PR}_{n}$. 

Finally, the delay in the wait state $D_{w_n}$ is calculated as the average time the sub-band becomes unavailable after the UL transmission, as a device moves from different states to the corresponding wait state in the $n^{th}$ transmission attempt.

\textbf{\underline{Energy vector}:} Similar to delay column vector, the energy in each state is given by the energy column vector, defined as [$E_{s_n}$, $E_{r^1_n}$, $E_{p^1_n}$, $E_{c^1_n}$, $E_{r^2_n}$,  $E_{p^2_n}$, $E_{c^2_n}$, $E_{w_n}$, $E_a$]. Here, $E_{s_n}$ is the energy spent by the end-device from the start of the $n^{th}$ UL frame transmission until it opens $RS_1$. $E_{r^1_n}$ and $E_{r^2_n}$ are the energy spent during the DL preamble reception time in $RS_1$ and $RS_2$ respectively, after the $n^{th}$ transmission attempt of the UL data frame. Consequently, $E_{c^2_n}$ is the energy consumed by the device during the rest of the ACK frame reception. Similar to $D_{c^1_n}$, $E_{c^1_n}$ consists of the energy used for receiving the ACK frame plus that spent in idle mode waiting for $RS_2$, after the respective UL transmission. From $D_{p^1_n}$, $D_{p^2_n}$ and $D_a$, the corresponding energy used for checking the preamble ($E_{p^1_n}$ and $E_{p^2_n}$), and after the detection of an ACK frame in the ACK state ($E_a$) will all be zero. $E_{w_n}$ is the energy consumed in the wait state when the device is idle, after using the sub-band during the $n^{th}$ transmission.
\section{Model Evaluation}
In this section, we show the delay and consumed energy per successful UL transmission results achieved through our analytical model, using a real network setup, as well as from the ns-3 based LoRaWAN module~\cite{IEEEhowto:09}. 
\subsection{LoRaWAN test-bed Experiments}
\subsubsection{Network Setup}
\begin{figure}[t!]
\begin{center}
\resizebox{3.1 in}{1.5 in}{\includegraphics{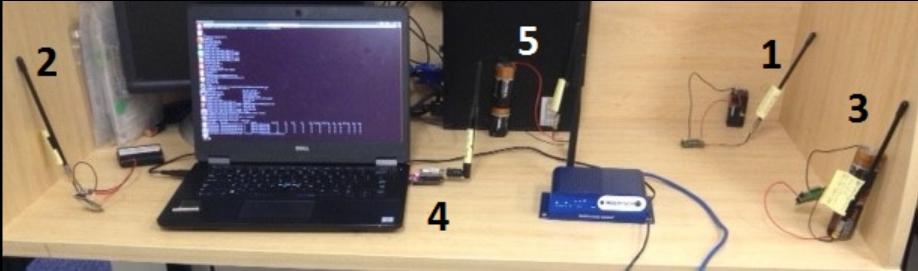}}
\caption{\em LoRaWAN real network setup with mDots and Multi-tech conduit.}
\label{jpg41}
\end{center}
\end{figure}

For a real network environment, we use a MultiConnect Conduit~\footnote{http://www.multitech.net} gateway and mDots as the end devices in its coverage area. Our setup (Fig.~\ref{jpg41}) shows the gateway with 5 devices used in our experiments. The LoRaWAN transmission parameters settings are the same as highlighted in Table~\ref{table2:parameters}. The mDot device performs over-the-air (OTA) activation to join the LoRaWAN network. A device records the time before each UL data frame transmission. Similarly, the time is also recorded immediately after the device receives the corresponding ACK frame. The difference of these two times gives the successful frame transmission to ACK delay. In the same way, the device maintains a transmission counter that shows the number of attempt needed until the successful transmission of the specific frame. An mDot is programmed to send the delay of each subsequent UL confirmed frame and the transmission counter value as a payload in the next data frame. Consequently, all UL frames from different network devices are collected at the network server, containing within their payloads the delay and transmission attempt information for each previous UL frame. 

Due to the limited number of available physical devices, we focus on a scenario where the UL duty cycle of each mDot is increased beyond the normal RDC limit (of 1\%). This is done to achieve a higher network load, and to evaluate the proposed analytical model as the network gets saturated. Hence, we increased the duty cycle from 1\% to 16\% in the presence of 5 devices as shown in Fig.~\ref{jpg41}. Thus 5 devices with 16\% duty cycle in real setup shows the performance that is comparable to 80 devices (each with 1\% duty cycle) using the proposed analytical model. For each duty cycle value, the setup runs over a duration of 2 hours.
\subsubsection{Evaluation Results}
\begin{figure}[!ht]
\begin{center}
\resizebox{3.0 in}{2.1 in}{\includegraphics{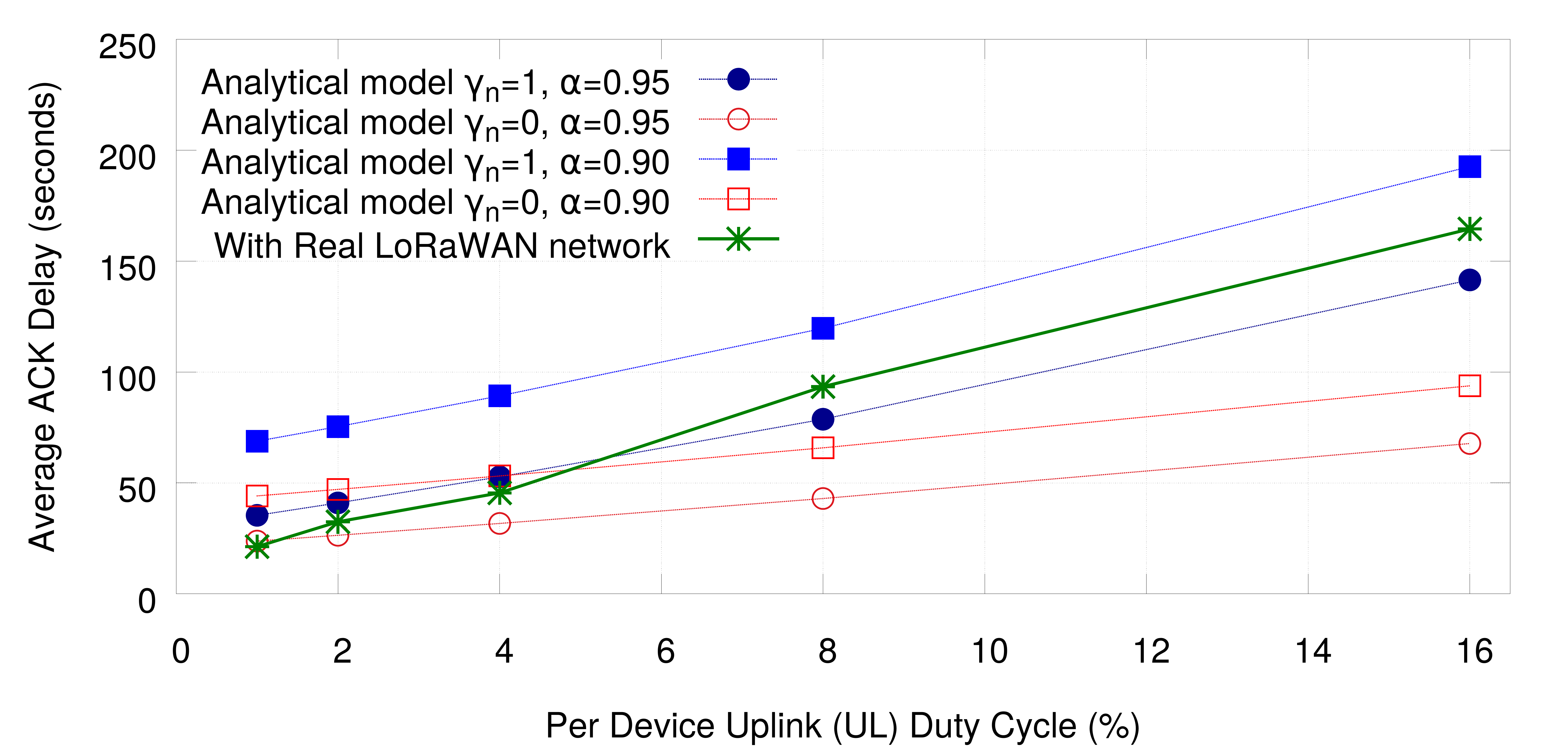}}
\caption{\em Analytical model evaluation with real network setup of A = 5, N = 8, $m_C$ = 7.}
\label{jpg411}
\end{center}
\end{figure}

The test-bed results of the average ACK delay are shown in Fig.~\ref{jpg411} with 5 devices with the increasing UL operation duty cycle per device. As we can see, with a larger duty cycle, the resulting delay increase becomes quite significant compared to the smaller duty cycle. The results from our approximated values via the analytical model achieved using the same network parameters with an almost perfect channel ($\alpha$=0.95) also show a similar trend with the increasing duty cycle. It is important to note that with higher duty cycle (replicating the operation of large number of network devices) a difference is observed between real network and our analytical model average ACK delay results. This is expected since with more simultaneous transmissions (larger duty cycle), accurate analytical estimation of channel conditions is non-trivial. Also note that from assumption 3 (in Section III A), the model does not consider the RDC limit over the reserved channels. Hence with larger duty cycle the network performance degrades which increases the overall observed average ACK delay.
\subsection{ns-3 Simulation Experiments}
\subsubsection{ns-3 LoRaWAN Module}
The ns-3 based LoRaWAN module is an extension of the ns-3 module for low power wide area network (LPWAN)~\cite{IEEEhowto:19}. The PHY layer of each device interacts with that of the respective gateway via the ns-3 Spectrum PHY module, implementing the device's air interface and channel specific parameters based on the path loss model. On each channel in a sub-band, a gateway can simultaneously receive signals with different spreading factors. Note that, in the module the LoRa PHY uses the error model drawn from the baseband implementation of the PHY layer in MATLAB based on an AWGN channel, as described in~\cite{IEEEhowto:09}.

We modified the functionality of the LoRaWAN module based on the LoRaWANv1.1 standard specifications. All end devices select a new channel among the available channels in a pseudo-random manner for each new UL frame transmission. In each retransmission, the device changes its channel by randomly picking a channel among the set of available channels, excluding the one used in the last transmission attempt. Then, after every two consecutive transmissions of an UL data frame (with the same counter value), the device reduces its rate by changing DR (/SF) by one step. 

The collision model used in the ns-3 LoRaWAN module considers both the impact of interference and the capture effect on the resulting bit error rate. Compared to this, our LoRaWAN analytical model uses a simple collision model based on simultaneous transmissions over the same channel. In order to account for the difference (due to capture effect) and reduce its impact, we restricted our evaluation experiments to scenarios with a relatively limited number of devices and with all devices in a distance of 5km from the gateway. While in our further evaluations (next section), the analytical model is used with even more number of devices.

Similar to real network, in the ns-3 LoRaWAN module, a network server can acknowledge a device in $RS_1$, following RDC restrictions provided that the MAC is in idle state. If this is not the case, the gateway will choose $RS_2$.
\subsubsection{Evaluation Settings and Environment}
\begin{table}[t!]
\centering
\caption{Default Parameters}
\label{table2:parameters}
\begin{tabular}{|c|c|}
\hline
Parameters   & Value \\
\hline
\texttt{$t_I$} & 1 \\
\hline
UL tx. power &  14dBm \\
\hline
Gateway coverage radius & 5km \\
\hline
Spreading factor (SF) & DR0/SF12 \\
\hline
Preamble length & 12.25 symbols (8 bytes) \\
\hline
Confirmed PHY Payload & 21 bytes \\
\hline
ACK payload & 12 bytes \\
\hline
Code rate  & $\frac{4}{7}$ \\
\hline
$\Delta$ & 1\% \\
\hline
$RS_1$ DR offset & 0 \\
\hline
Channel bandwidth & 125 kHz \\
\hline
Path loss model & LogDistancePropagationLoss \\
\hline
\end{tabular}
\end{table}
%
%
The evaluation of our analytical model uses the default parameters defined in Table~\ref{table2:parameters}, unless stated otherwise. For the calculation of the number of symbols, we assume that header compression is enabled while the data rate optimization feature is disabled. A perfect channel (i.e. $\alpha$=1) implying no channel losses is used in a network with a gateway and randomly distributed devices. There is no dwell time limit for the frame transmission duration (as in LoRaWAN specification~\cite{IEEEhowto:12}). Each UL data frame include 8 bytes of application payload, with 13 bytes PHY payload, which makes a total size of 21 bytes. The ACK payload frames from the gateway are considered without the optional FPort field, resulting in a size of 12 bytes. To make a reasonable comparison, no duty cycle restriction is imposed over the reserved channel, as assumed in our analytical model. An UL confirmed frame transmission is performed following a pseudo-random channel hopping pattern, with each time picking one of the 7 available channels uniformly randomly. The experimental results of ACK delay and consumed energy using ns-3 are collected from 100 simulation runs, and the average is calculated. For the consumed energy resources, we use the specification of the Semtech SX1272/73 device~\cite{IEEEhowto:20} with PA-Boost-ON state and a voltage of 1.5V. The transmit current at RFOP 17 dBm is 90mA. The idle (sleep) mode current (with transceiver off) is 0.1$\mu$A, the normal idle mode current is 1.5$\mu$A, and the receive mode current is 10.8mA.

\subsubsection{ns-3 Evaluation Results}
\begin{figure}[t!]
\begin{center}
\resizebox{3.5 in}{2.3 in}{\includegraphics{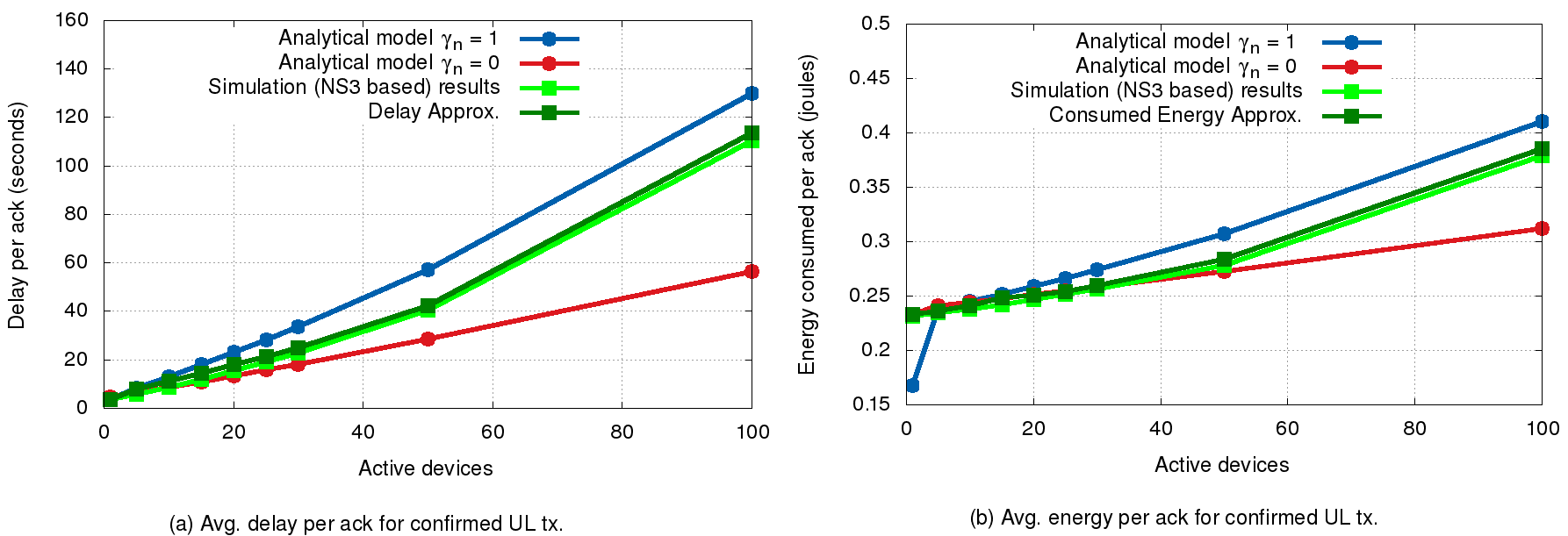}}
\caption{\em Analytical model evaluation of UL delay and energy via ns-3 simulations when $N$=8, $m_c$=7, and $\alpha$=1. }
\label{jpg1617}
\end{center}
\end{figure}
In this section, we use the system-level implementation of ns-3 based LoRaWAN module to evaluate the average delay and consumed energy per ACK'd UL frame results of the proposed analytical model for case 2, as defined in Fig.~\ref{jpg2}c. From Fig.~\ref{jpg1617}a, the results show that increasing the network load, by increasing the number of devices, significantly increases the average ACK delay for UL confirmed frame. Intuitively, the delay using $RS_1$ should be smaller compared to $RS_2$. However, in the worst case scenario the use of only $RS_1$ (when $\gamma_n$=1) for ACK transmission increases the delay and energy due to high interference from other transmissions and the possible collisions as the channel used in $RS_1$ is shared among all devices. In contrast to that, the channel used in $RS_2$ is reserved. Thus, as shown in Fig.~\ref{jpg1617}, the delay (/energy consumed) using $RS_2$ is small compared to $RS_1$, and it even surpasses the difference in their transmission times (of one second). Also, as with an increasing number of devices the collision probability increases, the difference in the delay and energy consumed using $RS_1$/$RS_2$ also increases. 

Furthermore, using our ns-3 based simulation experiments, we can see that the average ACK delay and energy confirms the validity of our analytical model. When the load is low, the ns-3 LoRaWAN results show a small delay and amount of energy consumed, as there are not many collisions, and ACKs can also be sent via $RS_2$. When the load becomes high, the average delay and energy consumed in ns-3 results increases, due to packet loss and retransmissions, caused by collisions and interference. Also due to the opportunistic use of $RS_2$ in our ns-3 LoRaWAN results, the resulting average delay and energy per successful ACK increases, but is still less compared to using only $RS_1$ ($\gamma_n$=1) slot for all ACK transmissions. 

From the results in Fig.~\ref{jpg1617} (a, b), it is important to note that the consumed resources depend heavily on the choice of receive slot (1 or 2). To get information about the choice of receive slot in each successive ($n^{th}$) transmission, we carried out multiple simulation runs in ns-3 with each set of devices (in Fig.~\ref{jpg1617}). The results give us the probabilities for the selection of the respective receive slots (1 and 2) by the gateway for ACK transmission, in response to a transmission attempt from a device. In other words, for each set of sub-band load we finds the probabilities of receive slots (1 and 2) selection in each of the $n^{th}$ transmission attempts. As expected, we found that with increasing load (network devices) the probability of receive slot selection in higher transmission attempts increases. We then used these probabilities of the choice of receive slot for ACK transmission and the expected delay (/energy) values from our analytical model to estimate the consumed resources when both slots (1 and 2) are available. The approximate results (represented as delay approx. and consumed energy approx.) in Fig.~\ref{jpg1617} shows that our model is able to accurately estimate the delay and consumed energy of LoRaWAN devices performing confirmed UL transmissions.

\section{Model Application Use Cases}
This section demonstrate the application of our analytical model in different network scenarios. Here, we consider a channel quality ($\alpha$) of 0.9. For the cases where these parameters are not varied, we use the maximum number of retransmissions, including the first transmission as $N=2$. Three channels are available for UL data transmission ($m_c$=3) in a network with total 50 devices ($A$=50). All other parameters are shown in Table~\ref{table2:parameters}. Further we evaluate the results using both cases 1 and 2, as discussed in Section~\ref{sec:back}.
\subsection{Number of Devices ($A$)}
\begin{figure}[t!]
\begin{center}
\resizebox{3.5 in}{2.1 in}{\includegraphics{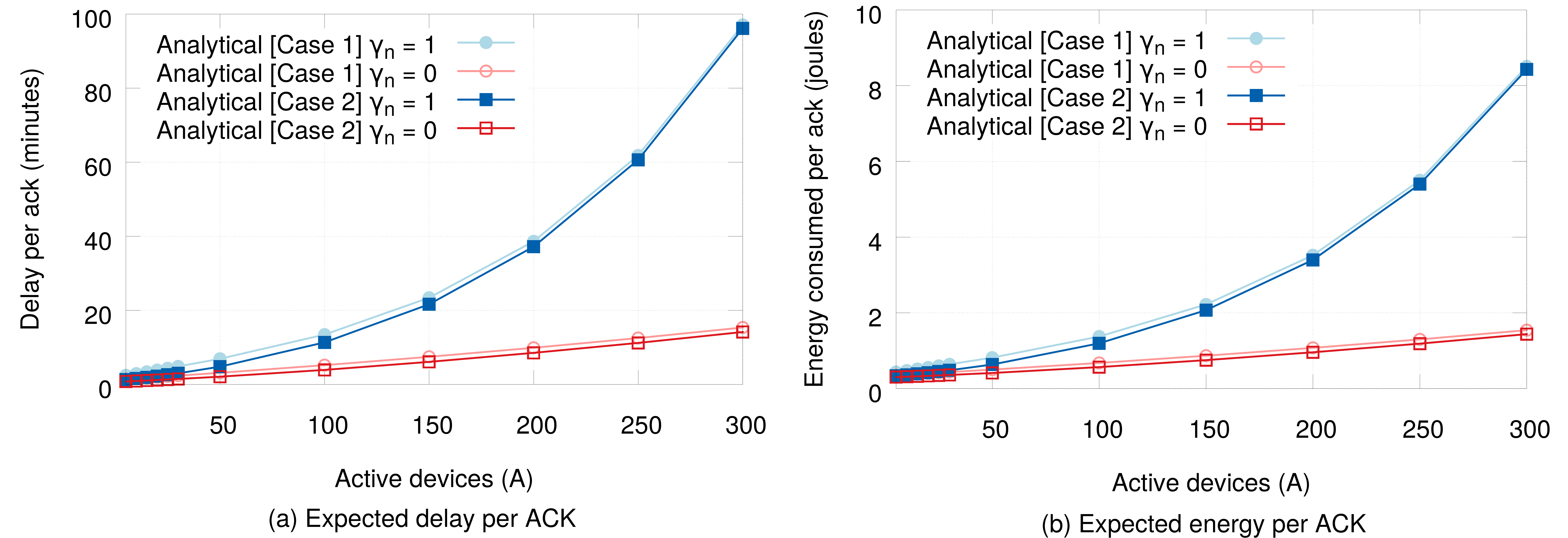}}
\caption{\em Expected delay and energy results per ACK'd transmission when $N$=2, $m_c$=3, and $\alpha$=0.9 for different values of $\gamma_n$ with respect to number of network devices.}
\label{jpg67}
\end{center}
\end{figure}

Results in Fig.~\ref{jpg67}a and Fig.~\ref{jpg67}b show the consumed resources with the increasing number of devices, i.e. load. As can be seen, increasing the number of devices also increases the resource consumption for an UL confirmed delivery. The trend in delay and consumed energy for the case of only $RS_1$ (i.e. $\gamma_n$=1) shows an exponential increase due to an increase in collisions. In contrast,  with the use of only $RS_2$ for ACK transmissions, the average delay and consumed energy increases only slightly with respect to the network load. This is because the reserved channel is only used by the gateway in $RS_2$, and the end device is not experiencing interference from the neighboring gateways. Hence, with high network load more devices can be served in a smaller amount of time, and the corresponding delay and energy per device for achieving a successful UL transmission is relatively small. Similarly, due to the small maximum transmission limit ($N$=2), a slight difference in delay and consumed energy results can be observed between the two cases with increasing network devices.
%
%
%
%
%
%
%
\subsection{Maximum Transmissions ($N$) Per Device}
\begin{figure}[t!]
\begin{center}
\resizebox{3.5 in}{2.2 in}{\includegraphics{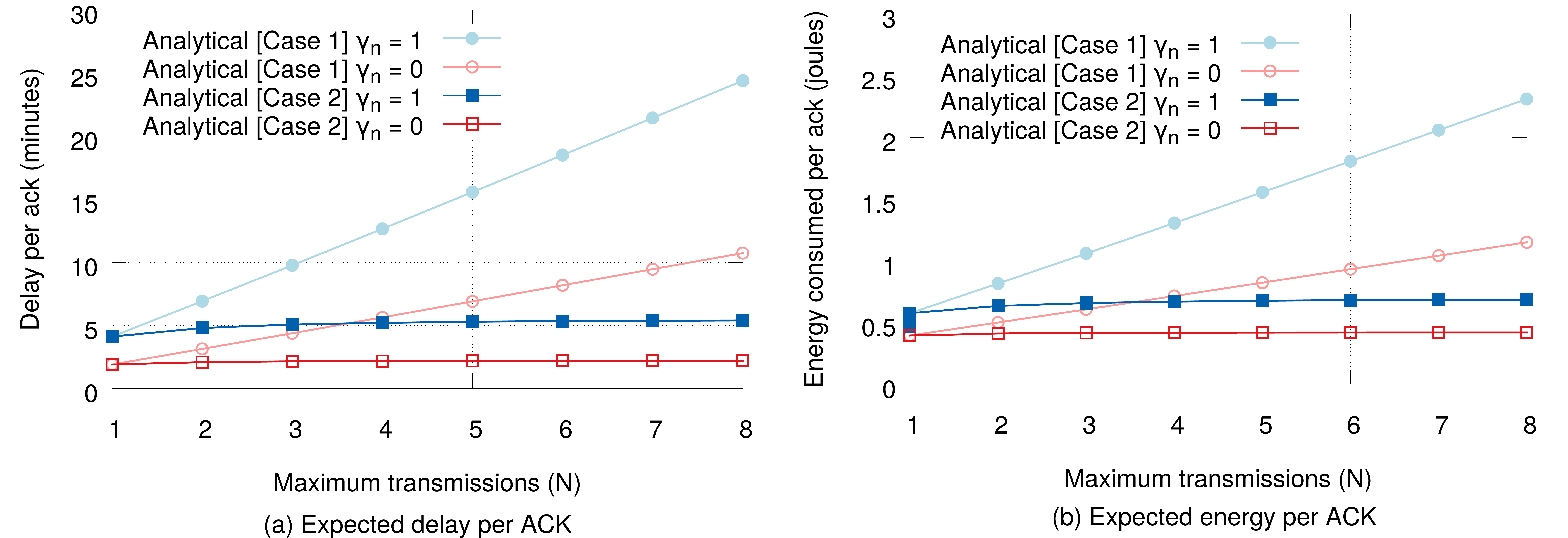}}
\caption{\em Expected delay and energy per ACK'd transmission when $A$=50, $m_c$=3, and $\alpha$=0.9 for different values of $\gamma_n$ with respect to maximum transmissions ($N$). }
\label{jpg1011}
\end{center}
\end{figure}

In the next scenario, we increase the maximum number of allowed transmissions $N$ of a frame, and evaluate the impact on the delay and consumed energy per ACK with 50 devices. Overall, in case 1 the results of Fig.~\ref{jpg1011}a and Fig.~\ref{jpg1011}b show an increasing trend of both delay and energy consumption with an increasing value of $N$, whereas the increase is minor for case 2. In case 1, when the ACK transmission is done using $RS_2$, we can see that the consumed resources increases with a smaller slope than using $RS_1$. This is because the use of reserved channel for DL ACK transmission eliminates the possibility of collisions. As a consequence, the probability of required retransmissions also reduces significantly, especially when the gateway is serving large number of devices. However, when only $RS_1$ is used to transmit ACK, the probability of a collision and subsequent retransmission increases. 

Finally, a considerable difference in average ACK delay and consumed energy can be noticed among the two cases (1 and 2), as the maximum transmissions ($N$) approaches 8. This further shows that when an ACK frame is lost, a higher limit non-uniformly increases the delay and consumed energy, especially when the ACK retransmission is not done.

\section{Related Work}
Recently, several efforts have been made to improve the performance of long range IoT technologies. The work in~\cite{IEEEhowto:03} gives a brief survey, comparing the LoRa and narrowband IoT (NBIoT) in terms of the cost, QoS, coverage, latency, and energy consumption. While the LoRa network can achieve low cost, prolonged battery life, and enhanced coverage, it does not provide any QoS. This is because the asynchronous nature of the LoRaWAN operation reduces the network server control over the delay and the data rate. An in-depth study and evaluation of LoRa and LoRaWAN technology is carried out in~\cite{IEEEhowto:08}. Using the physical LoRa testbed, the authors study the average throughput, collision rate, and channel utility in various network load conditions. The results show that with unconfirmed and confirmed transmissions the network achieves maximum channel occupancy when the load is 48\% and 25\%, respectively. 

The work presented in \cite{IEEEhowto:16} and~\cite{ IEEEhowto:08} shows that the RDC constraints limit the performance of Class A LoRaWAN devices. Especially for confirmed UL transmissions, this badly affects the LoRaWAN performance as the transmission of ACKs by the gateway increases the problem. Thus, as shown in~\cite{IEEEhowto:16}, the packet reception rate decreases due to a higher number of collisions when either the number of devices becomes too high, or their packet generation rate goes beyond a threshold. For these reasons, modifying the existing ALOHA-based channel access, investigating efficient channel hopping mechanisms, and devising new ways to minimize the number of ACKs emerge as new challenges in LoRaWAN.

The LoRaWAN performance highly depends on the appropriate selection of different transmission parameters (such as power, SF, coding rate, etc.) for each end-device. In~\cite{IEEEhowto:23}, the authors suggest a passive probing approach that chooses a suitable transmission power and UL rate to save device energy. The results shows that a significant amount of power can be saved using this method. However, it can take quite some time to collect enough probing measurements for an effective decision. Also, if the above parameters are not chosen well, it can increase the consumed energy, as well as a more delay due to the RDC limit. Therefore, an analytical model, such as presented in this paper, can be a useful tool for the exploration of LoRaWAN, and in particular the optimal choice of network and protocol parameters.

The works in~\cite{IEEEhowto:04},~\cite{IEEEhowto:05},~\cite{IEEEhowto:06}, and~\cite{IEEEhowto:07} model resource consumption for various aspects of LoRaWAN. In~\cite{IEEEhowto:04} and~\cite{IEEEhowto:07}, the authors present models to compute the energy and delay consumption for UL LoRaWAN transmissions. Specifically, the authors show how the frame collision rate and the respective device bit error rate (BER) changes under the given network conditions. In~\cite{IEEEhowto:04}, results show that a LoRa device with a 2400mAh battery can last up to 1 year when sending data frames every 5 minutes. The Markov-chain based queuing model presented in~\cite{IEEEhowto:07} is used to estimate the network transmission latency based on the traffic intensity. However, the model is not suitable for the case of retransmissions and does not consider confirmed devices~\cite{IEEEhowto:14} and~\cite{IEEEhowto:18}. 

The work presented in~\cite{IEEEhowto:05} models the behavior of the Class B DL transmissions using Markov-chain to calculate the data delivery delay to the end-device. The results show the delay with respect to load, available channels, and the used UL rates. As in the case of Class B devices, the gateway is continuously sending beacon frames (with 128s interval) for synchronization over the DL. The work considers that the gateway can only deliver the DL data either via ping slots or during one of the receive slots that opens after an UL transmission. Thus, the DL frame delay mainly depends on the number of ping-periods in a beacon interval. However, the work did not investigate how many ping periods per beacon interval are enough for efficient DL data delivery. 

One of the key contribution related to analytical modeling of LoRaWAN is presented in~\cite{IEEEhowto:06}. It models the resource consumption during the activation of devices in a network comprised of non-active and active devices. The results show that the activation delay and energy resources increases when more devices are going through the activation procedure. Similar to~\cite{IEEEhowto:05,IEEEhowto:06} our work is also based on a Markov-chain model, however it focuses on reliable transmissions and addresses the following limitations of previous works:
%
\begin{itemize}
\item It does not investigate both delay and energy consumption for confirmed UL transmissions in different cases.
\item It does not study retransmissions and their impact on the resource consumption and network performance.
\item None of the previous works evaluate results using both a real network and its system-level simulator.
\end{itemize}
\section{Conclusions}
This paper proposes and evaluates an analytical model for reliable UL transmissions in a LoRaWAN network. The model is able to quantify the resources required for achieving a successful transmission. Our test-bed and simulation based results show that the model can be use to estimate the delay and energy resources consumed in different use cases. Further the model also gives us the upper limit on the ACK delay and consumed energy when only $RS_1$ (i.e. $\gamma_n$=1) is use for ACK transmission. The proposed model can also be evaluated for different channel quality values, number of available channels, and the DR offsets used for sending ACKs in $RS_1$. 
%
%

%

\appendix[Formulation of State Transition Matrix]
This section explains the calculation of the state transition matrix~$\mathbf{P}$ for the proposed model shown in Fig.~\ref{jpg3}. As described in Section~\ref{sec:back}, once a device has sent its data frame, it will open the first receive slot after a RECV-DELAY time. Therefore, for the $n^{th}$ transmission of a frame the transition probability from state $s_n$ to $r^1_n$ will be $P_{s_n, r^1_n}$ $=$ 1 ($\forall$ n $\in$ [1, N]). In state ($R^1$, n), the device stays for the preamble air-time to receive the full preamble. The transition probability from state ($R^1$, n) to state ($P^1$, n) (i.e. $P_{r^1_n, p^1_n}$) is the probability that some data is received in receive slot 1 ($RS_1$). Next the transition from state ($P^1$, n) to state ($C^1$, n) occurs based on the probability that the detected preamble is correct, i.e. the preamble of the ACK sent by the gateway successfully arrives at the device in the $RS_1$. Furthermore, the probability of transitioning from state ($R^1$, n) to state ($R^2$, n), i.e. $P_{r^1_n,r^2_n}$, is the probability that nothing is detected in $RS_1$, as a result the device has opened $RS_2$. This is the probability that there was neither a transmission from the gateway nor from any of the devices during the $RS_1$.

\textbf{\underline{States for Receive Slot 1 ($RS_1$)}:} \textit{Preamble Detection State:} In the presence of $A$ devices, the probability that no UL transmission occurs is ${y_a}^{A}$. The probability that the ACK sent by the gateway is successful in $RS_1$ for the $n^{th}$ UL transmission attempt is $\alpha \gamma_n {y_a}^{A}$. Here, $\alpha$ represents the quality of UL channel used for frame transmission. The range of $\alpha$ is 0 $\leq$ $\alpha$ $\leq$ 1, with $\alpha$ $=$ 1 corresponds to a perfect channel, implying no frame loss occur due to the channel. $\gamma_n$ is the receive slot selection index for ACKs during the $n^{th}$ transmission attempt. $\gamma_n$ is $1$ when the first receive slot is used by the gateway for the ACK, while it is $0$ if the second receive slot is used for sending ACK. The probability that there is no transmission from the gateway is obtained by complementing the probability that the gateway transmits in $RS_1$ and is successfully received. We define the transition probability $P_{r^1_n,r^2_n}$ for $n=1$ as,
\begin{equation}\label{eq06}
	P_{r^1_1,r^2_1} = (1 - \alpha \cdot \gamma_1 \cdot {y_a}^{A}) \cdot {y_a}^{A},
\end{equation}
Similarly, the probability $P_{r^1_1, p^1_1}$ is defined as,
\begin{equation}\label{eq07}
	P_{r^1_1, p^1_1} = 1 - P_{r^1_1, r^2_1},
\end{equation}
Above findings (Eq.~\ref{eq06} and Eq.~\ref{eq07}) are true when n = 1. However when n $\geq$ 2, for $P_{r^1_n,r^2_n}$ and $P_{r^1_n, p^1_n}$, the $y_a$ is replaced by $y'_a$. 
%
%
%
%

\textit{Preamble Identification State:} Next, with a preamble of 8 bytes, the preamble detection duration (for 12.25 symbols) is 0.401408 sec which is less than one, even with the minimum rate (DR0/SF12). Therefore, $P_{p^1_n,w_n}$ $=$ $0$, as it can never be the case that the preamble detection takes more time than the time between two receive slots. $P_{p^1_n,c^1_n}$ is the probability that the preamble is correct, given that some data (/preamble) is detected. The probability that the detected preamble is correct in $RS_1$ is the probability that only one transmission happens either by gateway or by any one of the device in $RS_1$. The probability that only the gateway sends data (/preamble), and no other device is sending a data frame is $\alpha \gamma_1 {y_a}^{A} {y_a}^{A}$.  $( 1-\alpha \gamma_1 {y_a}^{A} ) {y_a}^{A - 1} x_a A$ is the probability that only one device has sent a frame, and the gateway has not sent any data over the DL. From the above findings, for $n=1$ we can define $P_{p^1_1,c^1_1}$ using the definition of conditional probability as follows,
\begin{equation}\label{eq11}
	P_{p^1_1,c^1_1} =\frac{( \alpha \gamma_1 {y_a}^{A} {y_a}^{A} ) + ( 1-\alpha \gamma_1 {y_a}^{A} ) {y_a}^{A - 1} x_a A}{P_{r^1_1,p^1_1}},
\end{equation}
Similarly, for n $\geq$ 2, $P_{p^1_n,c^1_n}$ can be find by replacing ${y_a}^{A}$ and $x_a$ in above by ${y'_a}^{A}$ and $x'_a$, respectively.
%
%
The probability of incorrect preamble detection, is the probability that the device move to the state ($R^2$, n) (i.e. $r^2_n$) from the state ($P^1$, n) (i.e. $p^1_n$) and is stated as,
\begin{equation}\label{eq12}
	P_{p^1_n,r^2_n} = 1 - P_{p^1_n,c^1_n},
\end{equation}
\textit{Checking Payload State:} For the first confirmed frame transmission, the probability that the received frame is successfully ACK'd in the first receive slot  ($P_{c^1_1,a}$), and no transmission from any device occurred is defined as,
\begin{equation}\label{eq13}
	P_{c^1_1,a} = \alpha \gamma_1 {y_a}^{A} \alpha {y_a}^{A},
\end{equation}
In the case of successive retransmissions ($n$ $\geq$ 2), ${y_a}^{A}$ in above equation will be replaced by ${y'_a}^{A}$.
%
%
Next the probability that for the first transmission of confirmed frame the corresponding payload with preamble does not contain any ACK is the probability that the frame gets corrupted due to channel errors therefore $P_{c^1_1,r^2_1}$ is defined as,
\begin{equation}\label{eq15}
	P_{c^1_1,r^2_1} = \beta_1 \alpha \gamma_1 {y_a}^{A} (1-\alpha) {y_a}^{A},
\end{equation}
Similarly, for the retransmission attempts (i.e. n $\geq$ 2), the transition probability $P_{c^1_n,r^2_n}$ can be calculated by replacing $\beta_1$ and ${y_a}^{A}$ in above equation by $\beta_n$ and ${y'_a}^{A}$, respectively.
%
%
In above, $\beta_n$ can be either 1 or 0, it is 1 if the ACK reception time in $n^{th}$ transmission ($t^{ACK}_{n}$) is less than one second otherwise it is 0. The probability $P_{c^1_n,w_n}$ can simply be stated as 1 - $P_{c^1_n,r^2_n}$ - $P_{c^1_n,a}$.

\textbf{\underline{Receive Slot 2 ($RS_2$) States}:} The probability that a UL frame is received correctly and the gateway is sending an ACK using $RS_2$ is defined below, 
\begin{equation}\label{eq17}
	P_{r^2_n,p^2_n} = \alpha (1-\gamma_n) {y_a}^{A},
\end{equation}
Next, the transition probability to the wait state, as a result of a failed reception in $RS_2$, is as follows,
\begin{equation}\label{eq18}
	P_{r^2_n,w_n} = 1 - P_{r^2_n,p^2_n},
\end{equation}
Since only the gateway is using the reserved channel for ACK transmission, $P_{p^2_n,c^2_n}$ $=$ 1, $P_{c^2_n,a}$ $=$ $\alpha$, and $P_{c^2_n,w_n}$ $=$ 1 - $\alpha$. Note that the above formulation is correct for the first transmission, while for retransmissions (when n = 2 ... N) the parameters $y_a$ and $x_a$ will be replaced by $y'_a$ and $x'_a$, respectively, as shown previously.

\textbf{\underline{Wait States ($w_n$) States}:} Next, $\forall$ n $\leq$ (N - 1), once the device PHY layer is in the wait state, the probability that it will go to the next send state is 1, $P_{w_n,s_{n+1}}$ = 1. Whereas upon n = N, the state transition probability to the next frame transmission becomes 1 (i.e. $P_{w_N,s_1}$ = 1), as $t_I$ is one. 

Above, we have defined the state transition matrix following the operation of LoRaWAN shown in Fig.~\ref{jpg2}c (case 2). The reader should note that during default LoRaWAN operation, as shown in Fig.~\ref{jpg2}b (case 1), once an ACK is sent, the network server should ignore all subsequent transmissions of the same frame. In such a case, the ACK from gateway at the $n^{th}$ (for n $\geq$ 2) transmission attempt of a device depends on the probability that the frame ACK has not been sent in any of the past n-1 transmission attempts. Thus, the probability that the gateway sends an ACK in response to the $n^{th}$ transmission attempt is denoted by $PR^{ACK}_{n}$ and defined as,
\begin{equation}\label{eq051}
	PR^{ACK}_{n} = [1 - (\alpha \cdot {y'_a}^{A} (\gamma_n \cdot {y'_a}^{A} + (1 - \gamma_n)))]^{n-1},
\end{equation}
where $\alpha {y'_a}^{A} \gamma_n {y'_a}^{A}$ and $\alpha {y'_a}^{A} (1 - \gamma_n)$ are the probabilities that an ACK is sent via $RS_1$ and $RS_2$, respectively. For case 1 (in Fig.~\ref{jpg2}b), the above state transition probabilities i.e. $P_{r^1_n,r^2_n}$, $P_{p^1_n,c^1_n}$, $P_{c^1_n,a}$, $P_{c^1_n,r^2_n}$, and $P_{r^2_n,p^2_n}$ are updated with $PR^{ACK}_{n}$ $\forall$ n $\geq$ 2.

\bibliographystyle{abbrv}

\end{document}